\documentclass[a4paper,12pt]{article}
\usepackage{epsfig}
\usepackage{amsmath}
\usepackage{amssymb}

\begin{document}
\def\beq{\begin{equation}}
\def\eeq{\end{equation}}
\def\bea{\begin{eqnarray}}
\def\eea{\end{eqnarray}}
\def\ve{\vert}
\def\nnb{\nonumber}
\def\ga{\left(}
\def\dr{\right)}
\def\aga{\left\{}
\def\adr{\right\}}
\def\rar{\rightarrow}
\def\nnb{\nonumber}
\def\la{\langle}
\def\ra{\rangle}
\def\ba{\begin{array}}
\def\ea{\end{array}}

\title{A study on the radiative dileptonic decays $B^0(B_s)\to
\gamma\ell^+\ell^-$\thanks{This work is partly supported by National
Science Foundation of China under Grant No.10475085, and Natural
Science Foundation of Hebei Province of China under Grant No.
A2005000535.}}

\author{Jun-Xiao Chen$^{a,b}\footnote{e-mail: chenjx@mail.ihep.ac.cn}$, Zhao-Yu Hou$^c$,
Cai-Dian L\"u$^a$\\
 {\it \small $a$   Institute of High Energy Physics,
CAS, P.O.Box 918(4)  Beijing 100049, China }\\
{\it \small $b$ College of Physics Science and Information
Engineering, Hebei Normal University,}\\ {\it \small  Shijiazhuang, Hebei 050016, China} \\
{\it \small$c$ Mathematics and Physics Department, Shijiazhuang
Railway Institute,  Shijiazhuang,} \\ {\it \small Hebei 050043,
China} }

\maketitle
\begin{abstract}
We study the rare radiative dileptonic decays $B^0(B_s)\to
\gamma\ell^+\ell^-$ ($\ell=e,\mu$)   in the standard model. By
using the $B$ meson wave function constrained by non-leptonic
decays, the branching ratios turn out to be of the order of
$10^{-9}$ for
 $B_s\to\gamma\mu^+\mu^-$, $\gamma e^+ e^-$ and $10^{-10}$ for
 $B^0\to\gamma\mu^+\mu^-$, $\gamma e^+ e^-$.
 Based on the study,  these decays are accessible
 at the near future LHC-b experiment, which are useful to determine
the $B(B_s)$ wave function.
\end{abstract}

\section{Introduction}{\label{sec:intro}}

The radiative and leptonic B decays
 have been the subject
of many theoretical studies in the framework of the Standard Model
(SM) and  the search of new physics. These processes play an
important role in determining the  parameters of the SM and some
hadronic parameters in QCD, such as the Cabibbo-Kobayashi-Maskawa
(CKM) matrix elements, the meson decay constant $f_{B}$, and
$f_{B_s}$. Especially, the radiative leptonic decays can provide
information on heavy meson wave functions \cite{li}. Being rare
decays in SM, they are also very sensitive to any new physics
contributions.

 Due to the GIM mechanism \cite{9704376}, there is absence of
flavor changing neutral current  transition at the tree level, thus,
the  pure leptonic processes  $B^0(B_s)\to \ell^+\ell^-$
($\ell=e,\mu$) can only occur via penguin and box diagrams. In
addition, rare decays of heavy pseudoscalar meson into light lepton
pairs are helicity suppressed. Some of the branching ratios are very
small, for instance \cite{2} $B_r(B_s\to\mu^+\mu^-)=2\times 10^{-9}$
and  $B_r(B_s\to e^+e^-)=4\times 10^{-14}$, so it is difficult for
us to extract the $B_s$ meson decay constant information from these
processes. For $B^0$ meson, the situation is even worse due to the
smaller CKM matrix elements. Although the process
$B_s\to\tau^+\tau^-$ is free from this  helicity suppression
mechanism, the branching ratio turns out to be around $8\times
10^{-7}$ in the SM \cite{3}, which is much larger than the branching
ratios of $B_s\to\mu^+\mu^-$, $B_s\to e^+e^-$, it is still hard to
be detected at   B-factory where the efficiency is not better than
$10^{-2}$.

 Because of massless   neutrino, the decay
$B^0(B_s)\to\nu \bar \nu$ is helicity forbidden. While, if an
additional real photon is emitted, the forbidden situation will
change \cite{9604378,0509093}. Similarly the helicity suppression
in the pure leptonic B decays $B^0(B_s)\to \ell^+\ell^-$ will be
cured in the radiative decay $B^0(B_s)\to \gamma \ell^+\ell^-$.
This is already shown in the simple constituent quark model
calculation \cite{9606444}, light cone QCD \cite{sum}.
   In this paper, we
employ the $B$ meson distribution amplitude derived from
non-leptonic B decays  to analyze these processes again.

We organize our paper as follows: In Sec.2, the relevant effective
Hamiltonian will be given in the SM. In Sec.3, the wave function of
$B$ meson will be used to calculate these processes, and later some
comparison will be given. Finally, Sec.4 includes a brief
conclusion.

\section{Effective Hamiltonian}\label{sec:hami}

\begin{figure}[htbp]
\begin{center}
\includegraphics [scale=1.0] {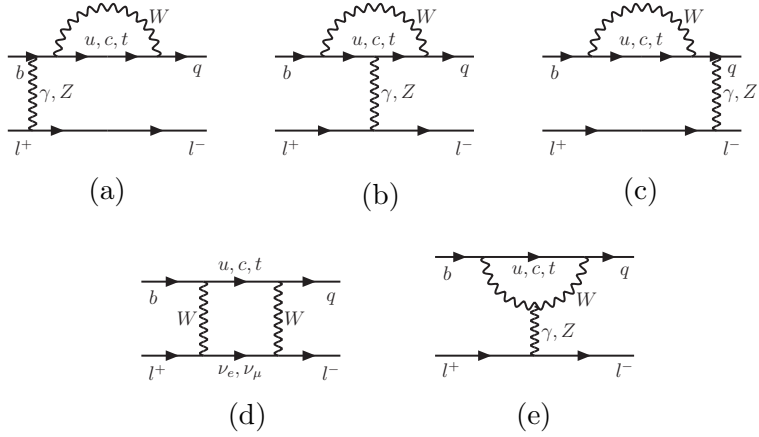}
\caption{Leading-order diagrams for $b \to q \ell^+\ell^-$
transition in SM.} \label{fig1}
\end{center}
\end{figure}

Let us start with the quark level processes $b\to q\ell^+\ell^-$,
with $q=s$ or $d$ quark, $\ell=e$ or $\mu$. The leading order
Feynman diagrams are shown in Fig.\ref{fig1}. It is easy to see that
the magnetic penguin, $Z$ penguin and box diagrams contribute to
these processes. They
  are subject to QCD corrections which can be obtained by
connecting the quark lines by   gluon lines. The effective
Hamiltonian in SM for them is \cite{5}:
\begin{eqnarray}
{\cal
H}_{eff}=\frac{\alpha{G_F}}{\sqrt{2}\pi}V_{tb}V^*_{tq}\left\{\left[\frac{2{C_7}{m_b}}
{q^2}\bar{q}P_R(\not \!p\gamma^\mu-\gamma^\mu\not\!p)b
+C^{eff}_9\bar{q}
\gamma^\mu{P_L}b\right]\bar{l}\gamma_\mu{l}+C_{10}(\bar{q}\gamma^\mu{P_L}b)~
\bar{l}\gamma_\mu\gamma_5{l}\right\},
\end{eqnarray}
where $P_L=(1-\gamma_5)/{2}$, $P_R=(1+\gamma_5)/{2}$,
$q^2=(P_++P_-)^2$, with $P_+$ and $P_-$ are the momenta of lepton
pair.
 $C_7$, $C^{eff}_9$, and $C_{10}$ are the $QCD$ corrected Wilson
 coefficients, whose specific forms are given in Ref.\cite{13}.

\begin{figure}[htbp]
\begin{center}
\includegraphics [scale=1.0] {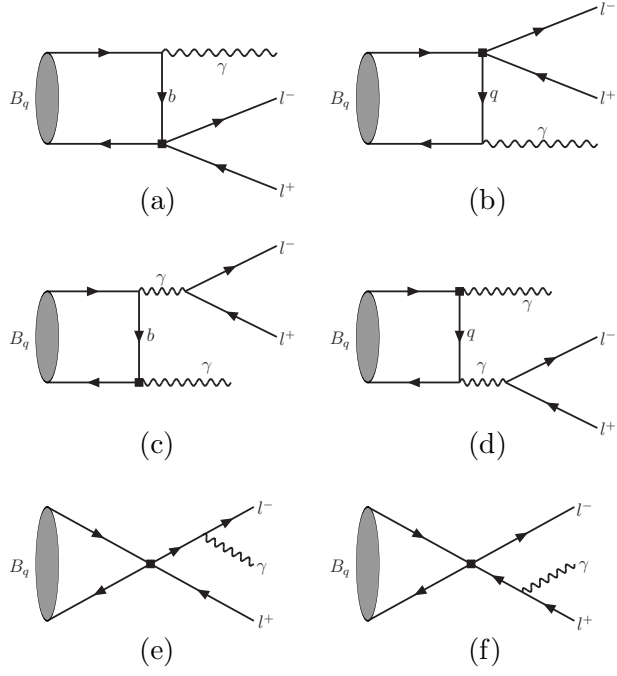}
\caption{Feynman diagrams for $B_q\to \gamma{\ell^+}{\ell^-}$ in
the SM (black dots are dimension 6 operators).} \label{fig2}
\end{center}
\end{figure}

If an additional photon line is attached to any of the charged lines
in Fig.\ref{fig1}, we will have the radiative leptonic decays $b\to
q \gamma \ell^+ \ell^-$. In fact, there are two kinds of diagrams:
photon connecting to the internal line of Fig.1, and photon
connecting to the external line of Fig.1. For the first kind of
diagrams, because of the effective operators are dimension-8 instead
of dimension-6, there will be a suppression factor of $m_b^2/m_W^2$
in the Wilson coefficients compared with the ones for $b \to
q\ell^+\ell^-$. Therefore we will only consider the second category
of diagrams. In this case, we only have dimension 6 operators with
an additional photon from any of the fermion lines, which are shown
in Fig.\ref{fig2}.

First, let us study the  Fig.\ref{fig2}(e) and (f) with photon
attached to external lepton lines.  Being a pseudoscalar meson,
 $B$ meson  can only  decay through axial current,  so
\begin{eqnarray}
<0|\bar{q}\sigma^{\mu\nu}{P_R}b|B>=0.
\end{eqnarray}
That is to say, in diagrams (e) and (f), the magnetic penguin
operator $O_7$ 's contribution vanishes. Our numerical calculation
shows that:
 the contribution
 from other operators for these two diagrams are also very small,
we can neglect their contribution safely. Similarly for
Fig.\ref{fig2} (c) and (d), We find numerically    they  are
 also negligibly small comparing with Fig.\ref{fig2} (a) and (b).
Thus the main contribution to $B^0(B_s)\to \gamma{\ell^+\ell^-}$
comes from the Fig.\ref{fig2} (a) and (b), which agrees with the
constituent quark model calculation \cite{9606444}.

Let us analyze the diagrams (a) and (b) in Fig.\ref{fig2}, with
photon emitted from the external quark lines $b$ or $q$. After
calculation, we get the amplitude:
\begin{eqnarray} \label{amp}
{\cal A} &=&\frac{e}{2}\bar{q}\left[\not\!\varepsilon
\frac{\not\!{k}-\not\!{p_2}+m_q}{p_2.k}\gamma^\mu{P_L}+
\gamma^\mu{P_L}\frac{\not\!{p_1}-\not\!{k}
+m_b}{p_1.k}\not\!\varepsilon\right
]b.\left[C^{eff}_9\bar{l^-}\gamma_\mu{l^+}+
C_{10}\bar{l^-}\gamma_\mu\gamma_{5}{l^+}\right]
\\
&+&\frac{e}{2}\frac{C_7m_b}{q^2}
\bar{q}\left[{P_R}(\not\!{q}\gamma^\mu-\gamma^\mu\not\!{q})\not\!
\frac{\not\!{p_1}-\not\!{k}+m_b}{p_1.k}\not\!\varepsilon
+\not\!\varepsilon\frac{\not\!{k}-\not\!{p_2}+m_q}{p_2.k}(\not\!{q}
\gamma^\mu-\gamma^\mu\not\!{q}){P_R}\right] b .[
\bar{l^-}\gamma_\mu{l^+} ],\nonumber
\end{eqnarray}
where $p_1$, $p_2$ and $k$ are the momenta of $b$, $q$ quark and
photon, respectively.

\section{Model calculations}\label{XXX}

 To simplify the  decay amplitude in eq.(\ref{amp}),
  we have to utilize the B meson wave function,
 which is not known from the first principal. Fortunately, many studies on non-leptonic
 B \cite{bdecay,cdepjc24121} and $B_s$ decays \cite{bs} have constrained the wave function
 strictly:
 \begin{equation}
\Phi_B= \frac{1}{\sqrt{6}} (\not \! p_B +m_B) \gamma_5 ~\phi_B
({x}), \label{bmeson}
\end{equation}
where the distribution amplitude $\phi_B$ can be expressed as
\cite{form}:
\begin{equation}
\phi_B(x) = N_B x^2(1-x)^2 \exp \left[ -\frac{M_B^2\ x^2}{2
\omega_b^2}  \right] \label{phib}.
\end{equation}
It satisfies the normalization relation:
\begin{equation}
\int_0^1\phi_B(x)dx=\frac{f_B}{2\sqrt{2N_c}}\;, \label{no}
\end{equation}
with $f_B$ is the $B$ meson decay constant.

 From definition of the wave function, we have $p_b=(1-x)P_B$, $p_q=xP_B$, and the decay
amplitude (\ref{amp}) is then:
\begin{eqnarray}
A=C\left[iC_1\epsilon_{\alpha\beta\mu\nu}P^\alpha_B\varepsilon^\beta{k^\nu}
+C_2(k_\mu\varepsilon_\nu-
\varepsilon_\mu{k}_\nu)P^{\nu}_{B_q}\right]
\left[(C^{eff}_9-\frac{2C_7M_B^2}{q^2})\bar{l^-}\gamma^\mu{l^+}+C_{10}\bar{l^-}
\gamma^\mu\gamma_{5}{l^+}\right]
\end{eqnarray}
with $C=\frac{e\alpha{G_F}}{\sqrt{12}\pi}|V_{tb}V^*_{tq}|,$  and
\begin{eqnarray}
C_1=\left( \int_0^1
\frac{\phi_B(x)}{1-x}dx+\int_0^1\frac{\phi_B(x)}{x}dx\right), \nonumber \\
C_2=\left( \int_0^1
\frac{\phi_B(x)}{1-x}dx-\int_0^1\frac{\phi_B(x)}{x}dx\right).\label{qmass}
\end{eqnarray}

After squaring the amplitude, and then performing the phase space
integration over one of the two Dalitz variables, we get the
differential decay width versus  the photon energy $E_\gamma$,
\begin{eqnarray}
\frac{d\Gamma}{dE_\gamma}=\frac{2\alpha^3{G^2_F}}{12^2\pi^4}|V_{tb}
V^*_{tq}|^2(C^2_1+C^2_2)(M_{B_q}-2E_\gamma) E_\gamma
\left[\left(C^{eff}_9-\frac{2M_{B_q}C_7}{M_{B_q}-2E_\gamma}\right)^2+C^2_{10}\right].
\end{eqnarray}

\begin{table}[htbp]
\caption{Comparison of branching ratios with other model
calculations }
\begin{center}
\begin{tabular}{|c|c|c|c|c|c|c|c|}
\hline\hline & Our Results  & Quark Model\cite{9606444} & light cone
 \cite{sum}\\
\hline
 $\mathbf{BR}(B_s\to \gamma\mu^+\mu^-)$
              & $ 1.7^{-0.46}_{+0.98}\times 10^{-9}$
              & $ 4.6 \times 10^{-9}$
              & $ 1.9 \times 10^{-9}$
             \\
\hline
 $\mathbf{BR}(B_s\to \gamma{e^+}{e^-})$
              & $1.9 ^{-0.52}_{+1.21}\times 10^{-9}$
              & $6.2\times 10^{-9}$
              & $2.35 \times 10^{-9}$
             \\
\hline
 $\mathbf{BR}(B^0\to \gamma\mu^+\mu^-)$
              & $0.65 ^{-0.23}_{+0.36}\times 10^{-10}$
              & $6.2\times 10^{-10}$
              & $1.2\times 10^{-10}$
            \\
\hline
 $\mathbf{BR}(B^0\to \gamma{e^+}{e^-})$
              & $0.83 ^{-0.21}_{+0.48}\times 10^{-10}$
              & $8.2\times 10^{-10}$
              & $1.5\times 10^{-10}$
             \\
\hline
\end{tabular}\label{table}
\end{center}
\end{table}

In numerical calculations, we use the following parameters
\cite{7}:
\begin{eqnarray}
&\alpha=\frac{1}{137}, ~ G_F=1.66\times 10^{-5}\mathrm{GeV}^{-2},
\omega_b=0.4,~ f_{B}=0.19~ \mathrm{GeV},
\nonumber\\
&\omega_{b_s}=0.5, ~f_{Bs}=0.24~ \mathrm{GeV},
|V_{tb}|=0.999, ~~|V_{td}|=0.007, ~~|V_{ts}|=0.041,\nonumber\\
& M_{B^0}=5.28~ \mathrm{GeV},~M_{B_s}=5.37~ \mathrm{GeV},~
\tau_B=1.54\times 10^{-12}s ,~ \tau_{B_s}=1.46\times 10^{-12}s.
\end{eqnarray}
After integration of phase space, we get the decay branching ratios
shown in Table \ref{table} together with numbers from other models.
 The input parameters
$\alpha$, $f_B$ and CKM factors will only give an overall factor to
the uncertainty of branching ratios,  which can be obtained easily,
thus we do not show them.  The uncertainty shown in the
Table~\ref{table} comes from the heavy meson wave function, by
varying the parameter $\omega_b=0.4\pm 0.1$, and
$\omega_{b_s}=0.5\pm 0.1$.
 From this strong sensitivity, we know that the radiative decays
 can provide information or constraints on the $B(B_s)$ meson wave
 functions. Surely more uncertainty can come from the next-to-leading order
 contribution which is difficult to estimate.
 From the table, we can see that our results are similar to the
results from the light cone sum rule \cite{sum}, but smaller than
the constituent quark model \cite{9606444}. If we interpret the
second term of eq.(\ref{qmass}) as the inverse of the constituent
quark mass, we will find that it corresponds to $m_d\simeq 473$MeV
which is larger than that used in ref.\cite{9606444}. The simple
picture of constituent quark model gives a larger result due to
the smaller constituent quark mass. But the much smaller result
for $B^0$ decays, which is even smaller than the light come sum
rule one \cite{sum}, is due to the smaller CKM factor $|V_{td}|$
used here.

\begin{figure}[htbp]
\includegraphics [scale=1.0] {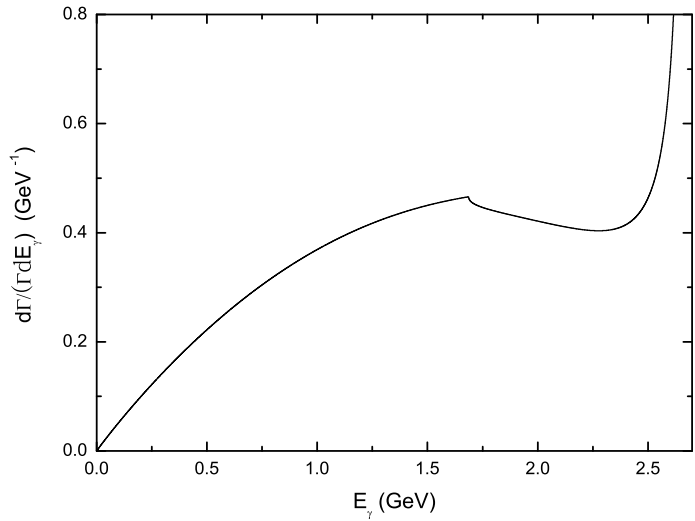}
\caption{Differential decay rate for $B_s (B^0)\to
\gamma{\ell^+}{\ell^-}$ versus the photon energy ${E_\gamma}$.}
\label{fig3}
\end{figure}

The differential decay width as a function of $E_\gamma$ is
displayed in Fig.\ref{fig3}. Most of the contribution is with an
energetic photon  which is easier for the experiment. One may
expect that through the mechanism of vector meson dominance
\cite{391461}, long distance effects also contribute to the
process $B^0(B_s)\to\gamma\ell^+\ell^-$ ($\ell=e,\mu$). Some
detailed calculation \cite{9702358} indicates that: the long
distance contributions are  significant in the resonance region.
It is probably not easy  to derive short distance information from
the dominant long distance contributions by total decay width.
Therefore, the photon energy spectrum shown in Fig.\ref{fig3} will
be helpful to distinguish those contributions.

\section{summary }\label{ccc}

We  calculate the rare decays $B^0(B_s)\to \gamma\ell^+\ell^-$\ in
SM. Utilizing the $B$ ($B_s$) meson wave functions constrained by
non-leptonic decays, the branching ratio is predicted  at the
order of $10^{-9}$ for $B_s\to \gamma\ell^+\ell^-$ and $10^{-10}$
for $B^0\to \gamma\ell^+\ell^-$. There are possibilities to detect
them in the LHC-b experiments, which could provide information on
the $B$ ($B_s$) meson wave function or new physics signal
\cite{new}.

\end{document}